\documentclass[twocolumn,reprint,epsf,graphics,psfig,superscriptaddress]{revtex4}

\usepackage{amsfonts}
\usepackage{graphicx}
\usepackage{amsmath}
\usepackage{amssymb}
\usepackage{latexsym}
\usepackage{psfrag}
\usepackage{dcolumn}
\usepackage{datetime}
\usepackage{multirow}
\usepackage{subfigure}
\usepackage{amssymb,amsmath,rotate,color}

\begin{document}
\title{Surface state transport in double-gated and magnetized topological insulators with hexagonal warping effects}
\author{Masomeh Arabikhah}
\affiliation{Department of Physics, Payame Noor University, P.O.
Box 19395-3697 Tehran, Iran}
\author{Alireza Saffarzadeh}
\altaffiliation{Author to whom correspondence should be addressed. Electronic mail: asaffarz@sfu.ca}
\affiliation{Department of Physics, Payame Noor University, P.O.
Box 19395-3697 Tehran, Iran} \affiliation{Department of Physics,
Simon Fraser University, Burnaby, British Columbia, Canada V5A
1S6}
\date{\today}

\begin{abstract}
We explore the scattering of Dirac electrons in a double-gated topological insulator in the presence of magnetic proximity effects and warped surface states.
It is found that a magnetic field can shift the Dirac cone in momentum space and deform the constant-energy contour, or opens up a band gap at the Dirac point, depending on the magnetization orientation. The double gate voltage induces quantum wells and/or quantum barriers on the surface of topological insulators, generating surface resonant tunnelling states. It is found that the hexagonal warping effect can increase the electronic transport at high energies when the constant-energy contour exhibits a snowflake shape. The energy-dependent conductances in the parallel and antiparallel magnetic configurations exhibit out-of-phase oscillations due to the quantum interference of propagating waves in the region between the two magnetized segments. Although the conductance spectrum of the double-well structure is higher than that of the double-barrier structure, the magnetoresistance ratio versus the separation distance between the two magnetized barriers exhibits pronounced oscillations due to the resonant tunnelling states. We show that the surface state transport can be controlled by the exchange field and gate voltage without breaking time reversal symmetry, suggesting that the double gated and magnetized topological insulators can be utilized to achieve a large magnetoresistance ratio with a tunable sign.

\end{abstract}
\maketitle

\section{Introduction}

Three-dimensional (3D) topological insulators are characterized by their gapless surface states and bulk gap as a consequence of time-reversal symmetry and band inversion induced by a strong spin-orbit coupling \cite{Hasan2010,Qi2011,Ando2013}. As a result of such a strong interaction, the magnetic heterostructures composed of topological materials have received significant attention for both their possible technological applications and the potential to promote our fundamental understanding of the underlying physics \cite{Han-PRL2017,Wang-NatComm2017,Yasuda-PRL2017}. Moreover, the spin–momentum locking in the surface states of the topological insulators is able to induce a non-equilibrium spin accumulation which can be electrically detected by measuring the hysteresis loops of the in-plane resistance of the magnetic tunnel junctions \cite{Li-NanoTec2014,Dankert-Nano-2015,Liu-Richardella-2015}.

On the other hand, gate voltages and external magnetic fields can cause many interesting effects in propagating behaviour of Dirac fermions by inducing electric and/or magnetic barriers on the surface of topological insulators \cite{Zhang-Wang-2012,Vali-Dideban-2016,Wang2012,Zhang-Zhai-2010,Song2014}. The electronic transport properties of a ferromagnet/normal/ferromagnet junction on the surface of a topological insulator showed that the conductance oscillates with the width of normal segment and gate voltage, like a spin field-effect transistor \cite{Zhang-Wang-2012}. Based on a transfer-matrix method, a theoretical investigation on the transport properties of Dirac electrons on the surface of 3D topological insulators under the modulation of electromagnetic superlattices has shown an imbalance between the number of transport channels for parallel and antiparallel magnetic configurations \cite{Wang2012,Zhang-Zhai-2010}. Moreover, the electric and magnetic barriers can reveal electron beam collimating property \cite{Wang2012}. Transport properties of electrons through a step junction, quantum wells and quantum superlattices on the surface of topological insulators have also suggested a clear oscillating behaviour in terms of incident angle of electrons, similar to those seen in Fabry–Perot interference in optics \cite{Song2014}. Chiba \textit{et al.}, \cite{Chiba-Takahashi-2017} modelled magnetic-proximity-induced magnetoresistance in disordered topological/ferromagnetic insulator bilayers by means of Kubo and Boltzmann theories. It was shown that for in-plane magnetizations the magnetoresistance ratio vanishes, while for out-of-plane magnetizations an energy gap opens up at the Dirac point which causes a large magnetoresistance value. Indeed, the interface between a ferromagnet and a topological insulator is considered as a spin source and accordingly, the spin-orbit coupling can enhance the magnitude of both charge and spin currents in the system \cite{Burkov-Hawthorn-2010,Ando-Hamasaki-2014}.

At low energies, the Fermi surface of the topological surface states is a circle and the electronic states near the Dirac point can be well described by the Dirac equation. As the Fermi energy increases, however, the shape of the constant-energy contour may change from a circle to a hexagon and then to a snowflake with sharp tips along the six  $\Gamma$-M directions. This phenomenon which is known as the hexagonal warping effect \cite{Fu2009,Nomura2014} and causes significant modifications in both DC conductivity \cite{Wang-Yu-2011} and optical conductivity \cite{Li-Carbotte-2013}, first observed in a Bi$_{2}$Te$_{3}$ sample \cite{Chen2009}. Angle-resolved-photoemission spectroscopy experiments have demonstrated that Bi$_{2}$Te$_{3}$ has a single Dirac cone on its surface \cite{Chen2009}. In addition to the in-plane spin polarization, the warping effect leads to an out-of-plane spin component which is carried by the surface states. 

Transport properties of topological insulators with warped surface states have recently been studied \cite{An2012,Li2013,Fu-Zhang-2014,Siu-Jalil-2014,Yu-Ma-2017,Akzyanov2018}. For instance, the scattering properties of a straight step defect on the surface of Bi$_{2}$Te$_{3}$ revealed a strong dependence on the direction in which the defect extends \cite{An2012}. At high energies where the warping effect is large, several critical momenta on the constant-energy contour are found, so that an incident wave with one of these momenta can be totally reflected or perfectly transmitted. There is always a finite reflection if the incident electron direction is not perfectly normal \cite{An2012}. Here, we would like to emphasize that not all the surface states of topological insulators exhibit the hexagonal warping effects. The size of the energy band gap and the crystal symmetry of topological insulators are crucial in their hexagonal warping strength of surface band dispersions \cite{Fu2009,Chen2009}. 

In this paper, we examine the influence of double-magnetized and double-gated regions on transport properties of Dirac fermions on the surface of a 3D topological insulator with hexagonal warping effect. We find that the resonant states and the electric conductance are strongly modulated by the gate voltage, incident energy and the exchange field strength. When the in-plane magnetizations are aligned in the growth direction, the time reversal symmetry is not broken and the conductance profile remains gapless for all electron energies. Although the conductance of double-well structure is nearly higher than that of the double-barrier structure, the magnetoresistance ratio versus the separation distance between the two magnetized regions exhibits pronounced oscillations due to the resonant tunnelling states. 

This paper is organized as follows: In Sec. II, we introduce our model and formalism for electron scattering from the double-gated and magnetized structure in the presence of hexagonal warping effect. By tuning our system parameters, numerical results and discussions for transmission properties, conductance spectra, and magnetoresistance ratios of incident electrons are presented in Sec. III. A brief conclusion is given in Sec. IV.
 \begin{figure}[hbt!]
\centerline{\includegraphics[width=0.95\linewidth]{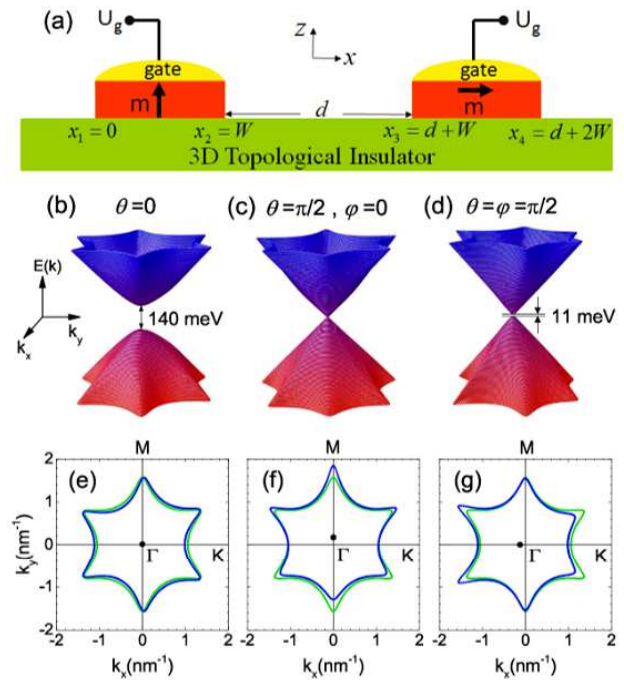}}
\caption{(a) Schematic view of a 3D topological insulator, modulated by double gate voltage and magnetization effect on its surface.
Panels (b), (c), and (d) show the Dirac cone of fermions on the surface of a 3D topological insulator in the presence proximity effect with magnetization orientation along the $z$, $x$, and $y$ axis, respectively. Panels (e), (f), and (g) show the constant-energy contours corresponding to Dirac cones (b), (c) and (d), respectively, at incident energy $E=400$\,meV. The black dots in (e)-(g) represent the Dirac point, while the green and blue solid lines are associated with $\Delta=0$, and $\Delta=70$\,meV, respectively. }
\label{F1}
\end{figure}

\section{Model and formalism}
We consider the 3D topological insulator $\mathrm{Bi}_{2}\mathrm{Te}_{3}$ which demonstrates strong warping effects and a Dirac cone on its surface. The interaction between bulk states and the surface states in such a structure can be ignored by tuning the Fermi level on only the surface states via appropriate doping \cite{Chen2009}. Therefore, we only focus on the topological surface states as one of the most important features of the topological insulators. We have applied gate voltages and magnetic proximity effects in two regions, called modulated regions of width $W$, separated by a distance $d$, as shown in Fig. 1(a). The effective Hamiltonian of the surface states in units of $\hbar=1$ and in the absence of particle-hole asymmetry can be given by \cite{An2012}:
\begin{equation}\label{eq1}
\hat{H}=\upsilon_{F}(k_{x}\sigma_{y}-k_{y}\sigma_{x})+\lambda(k_{x}^{3}-3k_{x}k_{y}^{2})\sigma_{z}+V(x) \ ,
\end{equation}
with
\begin{eqnarray}\label{eq11}
V(x)=(U_g&+&\Delta\mathrm{\bf{m}}\cdot\boldsymbol{\sigma})[\Theta(x)\Theta(W-x)\nonumber\\
&+&\Theta(x-d-W)\Theta(d+2W-x)]\ ,
\end{eqnarray}
where $\upsilon_{F}$ is the Fermi velocity, $\lambda$ is the warping parameter, and $\boldsymbol{\sigma}=(\sigma_{x},\sigma_{y},\sigma_{z})$ are Pauli matrices. $k_x=k\cos\beta$ and $k_y=k\sin\beta$ are the in-plane momentum components in which $\beta$ is the incident angle of electrons at $x=0$.
The first term in Eq. (\ref{eq1}) describes the helical Dirac fermions on the surface of the topological insulator (see Fig. 1a),
while the second term represents the hexagonal warping effect. $V(x)$ denotes the effects of gate voltage $U_g$ and magnetic proximity of
a ferromagnetic insulator with magnetization direction ${\bf m}=(\sin\theta\cos\varphi,\sin\theta\sin\varphi,\cos\theta)$ in the regions $0\le x\le W$ and
$d+W\le x\le d+2W$. Here,  $\theta$ is the polar angle with respect to the $z$-axis and $\varphi$ is the azimuthal angle measured from the $x$-axis. Also, $\Theta(x)$ is the step function and $\Delta$ is the proximity-induced exchange energy. Note that the gate voltage may act as a potential barrier $(U_g>0)$ or a quantum well $(U_g<0)$ for incident electrons with energy $E$.

The energy dispersion of the topological surface states in the presence of a gate voltage and magnetic proximity effects can be obtained by:
\begin{equation}\label{eq2}
E({\bf k})=U_g+s\epsilon({\bf k})\ ,
\end{equation}
with
\begin{equation}\label{eq3}
 \epsilon({\bf k})=\sqrt{[{{\lambda}({k_{x}}^{3}-3k_{x}{{k_{y}}^2})+	
\Delta\cos\theta}]^2+A}\, ,
\end{equation}
where $s={\pm 1}$ correspond to the upper and lower surface bands, respectively,
and $A=(\Delta\sin\theta\cos\varphi-\upsilon_{F}k_{y})^2+(\Delta\sin\theta\sin\varphi+\upsilon_{F}k_{x})^2$. We shall assume the Fermi level lies in the upper band $(s=+1)$.

Using $({k_{x}}^{3}-3{{k_{y}}^2}k_{x})=k^{3}\cos3\beta$, we can rewrite Eq. (\ref{eq3}) as
\begin{equation}\label{eq4}
\epsilon({\bf k})=\sqrt{\upsilon_{F}^2k^{2}+\Delta^{2}\sin^{2}{\theta}-B+C}\, ,
\end{equation}
where  $B=2\upsilon_{F}k\Delta\sin\theta\sin(\beta-\varphi)$ and $C=({\lambda}k^{3}\cos3\beta+\Delta\cos\theta)^2$.
If we ignore the warping effect and rewrite Eq. (\ref{eq4}) as a second-order equation in
terms of $k$, the condition of having real values for $k$ is obtained as
\begin{equation}\label{eq5}
 |E-U_g|\geq\Delta\sqrt{\sin^{2}\theta\cos^{2}(\beta-\varphi)+\cos^{2}\theta}.
 \end{equation}

In this case, when the magnetization direction is perpendicular to the surface of topological insulator ($\theta=\phi=0$), the largest energy gap, $2\Delta$, opens up at the Dirac point. This gap opening which suppresses the longitudinal conductivity can cause a large magnetoresistance ratio \cite{Chiba-Takahashi-2017}. For other magnetization directions, however, the energy gap is determined by $|2\Delta\cos\theta|$. In other words, the minimum value of the square root in Eq. (\ref{eq5}) can be obtained when $\cos(\beta-\phi)=0$. Therefore, in the case of in-plain magnetization ($\theta=\pi/2$), the energy dispersion becomes gapless.

In the presence of hexagonal warping effect, however, we find that the Fermi contour deforms into a snowflake shape. Moreover, the magnetic proximity effect can shift the Dirac cone and open up a gap in the band structure. In Figs. 1(b)-(d) we have shown the energy dispersion of the surface states when the magnetization is along the $z$, $x$, and $y$ axis, respectively. When the magnetization is fixed along the $z$ axis, the change in the Fermi contour is very small (see Fig. 1(e)), whereas the energy contour is modulated and shifted for the magnetization along $x$ and $y$ axes [see panels (f) and (g) in Fig. 1]. It is clear that in the case of magnetization along the $y$ axis, an energy gap which is much smaller than that with the magnetization along the $z$ axis opens up at the Dirac point. Nevertheless, since the gap width of 11 meV [panel (d) in Fig. 1] is much smaller than the thermal energy of 26 meV at room temperature, such a band gap does not considerably affect device performance.

Now we turn to electron transport through our main structure shown in Fig. 1(a). We consider an electron with momentum ${\bf q}=(q_{x},q_{y})$ and energy $E>0$ incident from left on the interface of normal/modulated regions at $x_1=0$. The energy and $y$-component of the momentum are conserved in the scattering process.
In the normal regions $(\Delta=0, U_g=0)$, the equation $E(q_{x},q_{y})=E_{F}$ in terms of $q_x$ is a sextic algebraic equation with six solutions.
If the energy is above the critical value $E_{c}\approx 377$ meV, the six roots are all real numbers, otherwise two roots are real and the rest are complex \cite{An2012}. In the modulated regions, however, the six roots are obtained from the equation $E(k_{x},k_{y})=E_{F}$ in terms of $k_x$.

By solving Eq. (\ref{eq1}), the corresponding eigenstates in the modulated regions can be written as:
\begin{equation}\label{eq7}
u({\bf k}_{m},{\bf r})=\sqrt{\frac{|\varphi_{1m}|^2}{|\varphi_{1m}|^2+|\varphi_{2m}|^2}}
\left(\begin{array}{cc}
1 \\  \frac{\varphi_{2m}}{\varphi_{1m}}
\end{array}\right)e^{ik_{x,m}x}e^{ik_{y}y}\ ,
\end{equation}
with
\begin{equation}\label{eq8}
\varphi_{1m}(k_{x,m},k_{y})=\Delta\cos\theta+\epsilon({\bf k})+\lambda k_{x,m}(k_{x,m}^{2}-3k_{y}^{2})\, ,
\end{equation}
\begin{equation}
\varphi_{2m}(k_{x,m},k_{y})=\Delta\sin\theta\,e^{i\phi} - \upsilon_{F}(k_{y}-ik_{x,m})\, ,
\end{equation}
where $m=1-6$. Moreover, the electron eigenstates $u_{m}({\bf q},{\bf r})$  in the normal regions follow similarly from Eq. (\ref{eq7}).
As a result, for an incident electron with given energy $E_{F}$, the wave functions of the scattering states in all regions can be written as
\begin{align}
\psi_{1}({\bf r})&=u(q_{x}^{i},q_{y})+\sum_{m=1}^{3} r_{m}u(q_{x,m}^{r},q_{y})~~~~x\le 0\ ,\\
\psi_{2}({\bf r})&=\sum_{m=1}^{6} s_{m}u(k_{x,m},k_{y})~~~~~~0\le x\le W \ ,
\end{align}
\begin{multline}
\psi_{3}({\bf r})=\sum_{m=1}^{3} f_{m}u(q_{x,m}^{r},q_{y})+\sum_{m=1}^{3} g_{m}u(q_{x,m}^{t},q_{y}) \\
W\le x\le d+W \ ,
\end{multline}
\begin{align}
\psi_{4}({\bf r})&=\sum_{m=1}^{6} h_{m}u(k_{x,m},k_{y})~~~~d+W\le x\le d+2W\, ,\\
\psi_{5}({\bf r})&=\sum_{m=1}^{3} t_{m}u(q_{x,m}^{t},q_{y})~~~~x\ge d+2W\, ,
\end{align}

Here $u(q_{x}^{i},q_{y})$ is the incident eigenstate, $g_{m}$ and $t_{m}$ are the transmission amplitudes with eigenstates $u(q_{x,m}^{t},q_{y})$ in the central and right normal regions, respectively, $r_{m}$ and $f_{m}$ are the reflection amplitudes associated with the eigenstates $u(q_{x,m}^{r},q_{y})$ in the left and central normal regions, while
 $s_{m}$ and $h_{m}$ are the scattering amplitudes in the modulated regions.

For fixed value of Fermi energy $E_F$ and $q_y$, the solution of $E(q_{x},q_{y})=E_{F}$ is dependent on whether $E_F<E_c$ or $E_F>E_c$. Therefore, there are two types of solutions. The first type which corresponds to $E_F<E_c$ consists of two real roots and four complex ones. Therefore, since the incident wave vector $q^i$ must be a real number, the conditions for the reflected wave vector $q^r$ and the transmitted wave vector $q^t$ can be chosen as $q^{i}_x=q^{t}_{x,1}=-q^{r}_{x,1}$ (real roots) and $q^{t}_{x,2(3)}=-q^{r}_{x,2(3)}$ (imaginary roots) \cite{An2012}. For the second type of solutions corresponding to $E_F>E_c$, all the roots are real numbers (three positive roots and three negative roots with the same absolute values) so that, two of them are hole-like propagating waves $(q_x\upsilon_{x}({\bf q})<0)$ and the other roots are electron-like waves $(q_x\upsilon_{x}({\bf q})>0)$. Considering the fact that the incident wave vector must be real, the wave vectors can be chosen as $q^{i}_x=q^{t}_{x,1}=-q^{r}_{x,1}$, $q^{t}_{x,2(3)}=-q^{r}_{x,2(3)}$, $q^{t}_{x,1(2)}>0$,  $q^{t}_{x,3}<0$, $\upsilon_ {x}(q^{t}_{x,m},q_{y})>0$ \cite{An2012}. Here, $\upsilon_{x}(q)=\frac{\partial E}{\partial q_{x}}$ is the electron group velocity along the $x$ axis.

Since the Hamiltonian Eq. (\ref{eq1}) is a third-order partial differential equation with respect to $k_{x}$, to determine the reflection and transmission amplitudes, the following boundary conditions should be satisfied:
\begin{equation}\label{eq10}
\begin{array}{ccc}
\psi_{n}({\bf r})\mid_{r=(x_{n},y)} = \psi_{n+1}({\bf r})\mid_{r=(x_{n},y)} \, , \\\\
\partial_{x} \psi_{n}({\bf r})\mid_{r=(x_{n},y)} = \partial_{x}\psi_{n+1}({\bf r})\mid_{r=(x_{n},y)} \, ,  \\\\
\partial_{x}^{2}\psi_{n}({\bf r})\mid_{r=(x_{n},y)} = \partial_{x}^{2}\psi_{n+1}({\bf r})\mid_{r=(x_{n},y)} \, ,
\end{array}
\end{equation}
with $n$=1-4. Using the above boundary conditions and the transfer matrix method which connects the incident wave to the
transmitted wave \cite{Saffar2003,Saffar2005,Li2013},  all the reflection and transmission amplitudes can be determined.
\begin{figure}
\centerline{\includegraphics[scale=0.58]{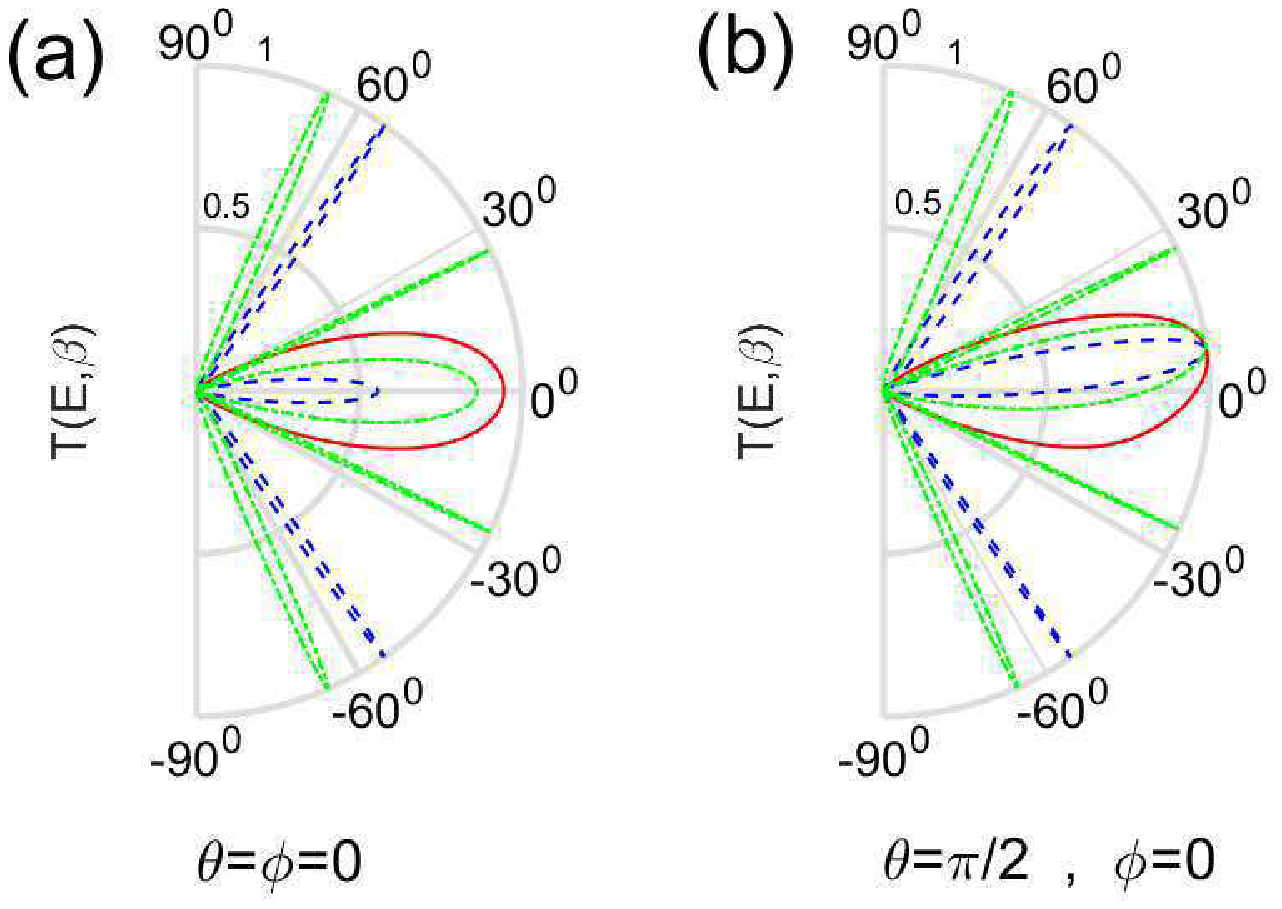}}
\caption{Transmission probability $T$ as a function of the incident angle $\beta$ with magnetization angles (a) $\theta=\varphi=0$, and
 (b) $\theta=\pi/2, \phi=0$ at different energies $E=200$ meV (red), $E=300$ meV (blue), and  $E=400$ meV (green).
The other parameters are  $U_{g}=300$ meV, $\Delta=30$\,meV and $d=W=4$\,nm. }
\label{F2}
\end{figure}

\begin{figure}
\centerline{\includegraphics[scale=0.54]{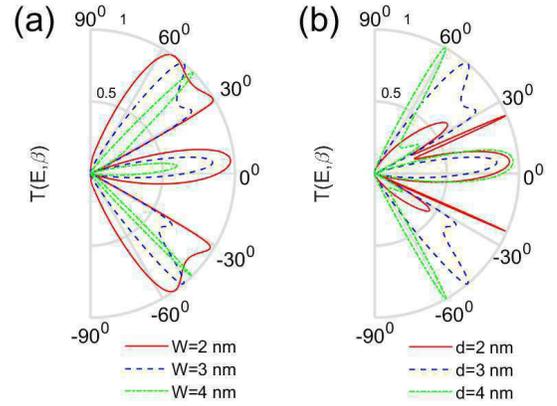}}
\caption{Transmission probability $T$ as a function of the incident angle $\beta$ with $E=U_{g}=350$ meV, $\Delta=50$ meV and $\theta=\varphi=\pi/4$ for  (a) $d=3$ nm and (b) $W=3$ nm.}
\label{F3}
\end{figure}

For the first type of solution, the total transmission coefficient of electrons with energy $E$ and incident angle $\beta=\arctan(\frac{q_{y}}{q_{x}})$ is given by $T(E,\beta)=|t_1|^2$, whereas for the second type of solution, the total transmission coefficient is calculated by $T(E,\beta)=\sum_{i=1}^3|t_i|^2$. Moreover, using the continuity equation $\nabla\cdot\mathbf{J}+\frac{\partial\rho}{\partial t}=0$, the charge current density $\mathbf{J}$ of an electron with charge $e$ can be expressed by the following components:
\begin{equation}\label{jx}
J_x=ev_F(\psi_l^\dagger\sigma_y\psi_l) + 3e\lambda k^2\cos 2\beta(\psi_l^\dagger\sigma_z\psi_l)\ ,
\end{equation}
\begin{equation}
J_y=-ev_F(\psi_l^\dagger\sigma_x\psi_l) - 3e\lambda k^2\sin 2\beta(\psi_l^\dagger\sigma_z\psi_l)\ ,
\end{equation}
where $\rho$ is the charge density and $k$ is the magnitude of wave vector in region $l (\equiv1-5)$. Therefore, assuming that the system has a large width in the $y$ direction given by $L_y$, the net current going from left to right is given by
\begin{equation}
I=\frac{L_y}{2\pi^2}\int_0^\infty\int_{-\pi/2}^{\pi/2} J_x [f_1(\epsilon_\mathbf{k})-f_5(\epsilon_\mathbf{k})]kdkd\beta\  ,
\end{equation}
where $f_l(\epsilon)$ is the Fermi function in region $l$. Considering Eq. (\ref{jx}) for region 5 and the difference of Fermi functions at low temperature and a small bias voltage, the differential conductance $G$ of electrons at Fermi energy can be obtained by
\begin{equation}\label{G}
G=g_0\int_{-\pi/2}^{\pi/2} T(E_F,\beta)F(k_F(\beta),\beta) d\beta\  ,
\end{equation}
with
\begin{equation}
F(k_F,\beta)=\frac{[2Bv_F^2k_F^{2}\cos\beta+3\lambda k_F^3(B^2-v_F^2k_F^2)\cos 2\beta]v_{F}}{E_{F}(B^2+v_F^2k_F^2)[\frac{dE(k,\beta)}{dk}]|_{k_F(\beta)}}\  ,
\end{equation}
\begin{equation}
B=\lambda k_F^3\cos 3\beta+\sqrt{v_F^2k_F^2+\lambda^2 k_F^6\cos^2 3\beta}\  ,
\end{equation}
\begin{equation}
k_F=\sqrt{[\sqrt{\frac{\kappa_{2}^3}{27}+\frac{\kappa_{1}^2}{4}}+\frac{\kappa_{1}}{2}]^{1/3} - [\sqrt{\frac{\kappa_{2}^3}{27}+\frac{\kappa_{1}^2}{4}}-\frac{\kappa_{1}}{2}]^{1/3}}\  ,
\end{equation}
where $k_F\equiv k_F(E_F,\beta)$ is the Fermi wave vector of incident electrons, $\kappa_{1}=\frac{E^2_F}{\lambda^2\cos^2 3\beta}$,
$\kappa_{2}=\frac{v_F^2}{\lambda^2\cos^2 3\beta}$, and $g_0=\frac{e^2E_FL_y}{2\pi^2v_F}$ is the unit of conductance with $\hbar=1$.

Note that, we expect the maximum value of conductance to be $2g_0$ due to the integral over $\beta\in [-\frac{\pi}{2},\frac{\pi}{2}]$. Such an interval for angle of incidence is especially necessary when the mirror symmetry in the constant-energy contour is broken in the presence of an external  magnetic field (see Fig. 1(f)).

\section{Results and discussion}
We now use the above method to obtain the numerical results of electronic transport through the double modulated structure presented in Fig. 1(a). We choose $\upsilon_{F}=2.55$ eV\,$\mathrm{\AA}$ and $\lambda=250$ eV\,$\mathrm{\AA^3}$ which produce the Fermi surface of $\mathrm{Bi}_{2}\mathrm{Te}_{3}$ in good agreement with experiments \cite{Chen2009}. The electron scattering in our system is governed by the exchange field strength $\Delta$, applied gate voltage $U_g$, the width of gate regions $W$, the separation distance $d$ and the magnetic alignments of the two modulated regions.

In Figs. 2(a) and (b), we have depicted the variation of transmission probability with incident angle of electrons through the structure, when ${\bf m}$ in both modulated regions is aligned along the $z$- and $x$-directions, respectively. As can be seen, the the system is fully transparent ($T=1$) at some incident angles, whereas the normally incident electrons are not perfectly transmitted due to the induced magnetic field. Indeed, the perfect transmission found for normally incident Dirac fermions is shifted from normal incidence ($\beta=0$) to an off-normal angle ($\beta\neq 0$) when the  magnetic fields align along the $x$-direction (see Fig. 2(b)) \cite{Katsnelson2006,Li2015}. Such an effect, however, does not occur for the case of ${\bf m}$ in the $z$-direction, as a result of broken time-reversal symmetry. Note that time reversal symmetry on the surface of each modulated region is dependent on the magnetization orientation in that region. Therefore, the time reversal symmetry remains broken as long as the magnetization orientations in both modulated regions are not fully aligned in the $x$-direction.
\begin{figure}
\centerline{\includegraphics[scale=0.55]{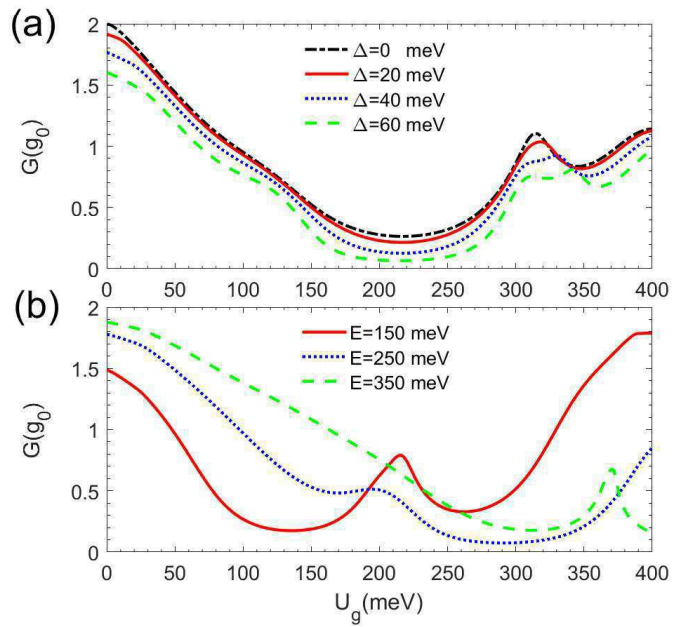}}
\caption{Calculated conductance as a function of gate voltage $U_{g}$ (a) at different values of $\Delta$ but fixed $E=$\,200 meV and (b) various incident energies $E$ but fixed $\Delta=50$\, meV. The other parameters are $\theta=\phi=\pi/4$ and $W=d=5$\,nm.}
\label{F4}
\end{figure}

The number of tunnelling resonances increases with increasing the incident electron energy and that the incident angles at which the resonances occur are considerably energy dependent. When $\bf m$ aligns along the $z$-direction the transmission profile is symmetric about $\beta=0$ while, this symmetry is broken for the case of $\bf m$ in the $x$-direction. This effect is fully associated with the broken mirror symmetry in the constant-energy contour around $\Gamma$-K direction as shown in Fig. 1(f). For the special case of $E=U_{g}=300$\,meV with $\bf m$ in the $z$-direction a strong suppression in the transmission probability of normally incident electrons happens compared to the transmission panels in the case of $E\neq{U_{g}}$. Due to the gap opening in energy dispersion and the influence of warping effect (see Fig. 1(b)), this behaviour is in contrast to p-n-p graphene junctions, where the transmission is nonzero and perfect ($T=1$) only at normal incidence \cite{Li2015}.
\begin{figure}
\centerline{\includegraphics[scale=0.68]{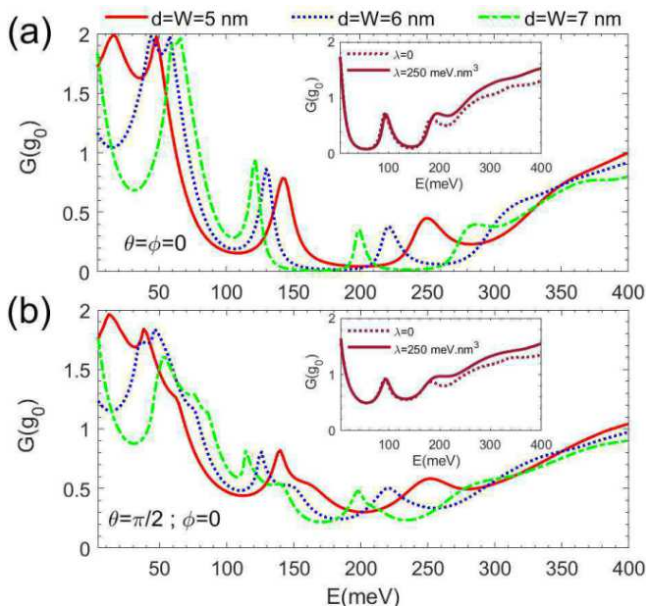}}
\caption{Calculated conductance as a function of incident energy $E$ with $U_{g}=200$\,meV, $\Delta=50$\,meV
for parallel magnetization directions at (a) $\theta=\varphi=0$, and (b) $\theta=\pi/2$, $\phi=0$. The insets represent the conductance vs incident energy in the presence and absence of warping effect with $U_{g}=100$\,meV and $d=W=5$ nm.}
\label{F5}
\end{figure}

In Figs. 3(a) and 3(b), we show our analysis of the transmission probability for $\theta=\phi=\frac{\pi}{4}$
with different $W$ values but fixed separation distance $d$ and with different $d$ values but fixed $W$, respectively.
It can be seen that with decreasing $W$, the transmission lobes become progressively wider and that the $T(E,\beta)$ profile approaches the maximum value around normal incidence, confirming the role of proximity effect in suppression of transmission for normal incident electrons.
Moreover, as the spatial separation $d$ between the two modulated regions is increased the transmission probability spans over a wider angular range.
For a fixed magnetization direction, the transmission probabilities are equally shifted from the normal incidence, regardless of the values of $W$ and $d$, as shown in Fig. \ref{F3}.
\begin{figure}[h]
\centerline{\includegraphics[scale=0.55]{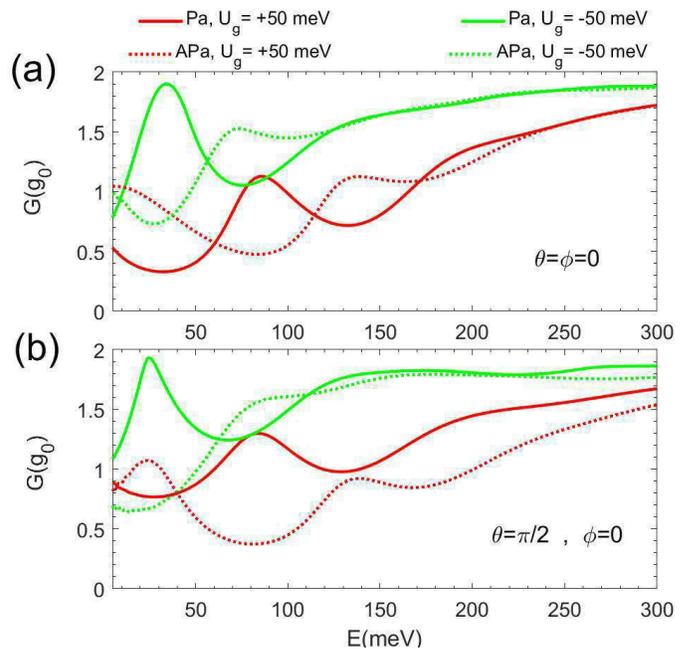}}
\caption{Calculated conductance as a function of incident energy $E$ with $\Delta=40$\,meV and $d=W=4$\,nm,
when the magnetization orientations align along the (a) $z$- and (b) $x$-axes.}
\label{F6}
\end{figure}

To investigate the effect of double gate voltage on transmitted electrons, we have depicted in Fig. 4(a) the conductance versus gate voltage for fixed parallel magnetization directions with $\theta=\phi=\frac{\pi}{4}$ at different exchange energies $\Delta$. In such a magnetic direction an energy gap opens up around the Dirac point as a result of nonzero $\bf m$ component in $z$-direction. We see that the conductance decreases with increasing the exchange field and hence the band gap (Fig. 4(a)). For instance, in the case of $\Delta=60$ meV, we found an energy gap of 86.4\,meV around the Dirac point and the conductance reaches a minimum and nearly constant value in the interval $180<U_g<250$\,meV around the gate value at which $U_g=E=200$\,meV. As $U_{g}$ increases, $G$ decreases for all $\Delta$ values until the Dirac point approaches the incident energy at which the coherent tunnelling governs the mechanism of charge transport. Then the conductance gradually increases with gate voltage to almost 1$g_0$ and exhibits distinct resonant peaks. The magnetic proximity effect reduces the conductance in the system, while the behaviour of $G$ remains nearly unchanged with gate voltage. Such a behaviour has also been seen in p-n-p graphene heterojunctions under the influence of external magnetic field \cite{Li2015}.
Moreover, Fig. 4(b) shows the conductance as a function of gate voltage at different incident energies $E$. At $U_{g}<E$, the conductance has larger values for higher energies and its minimum value is shifted along the voltage axis as the energy is increased. The position and the hight of resonant peaks are dependent on $\Delta$ and $E$ values.
\begin{figure}
\centerline{\includegraphics[scale=0.55]{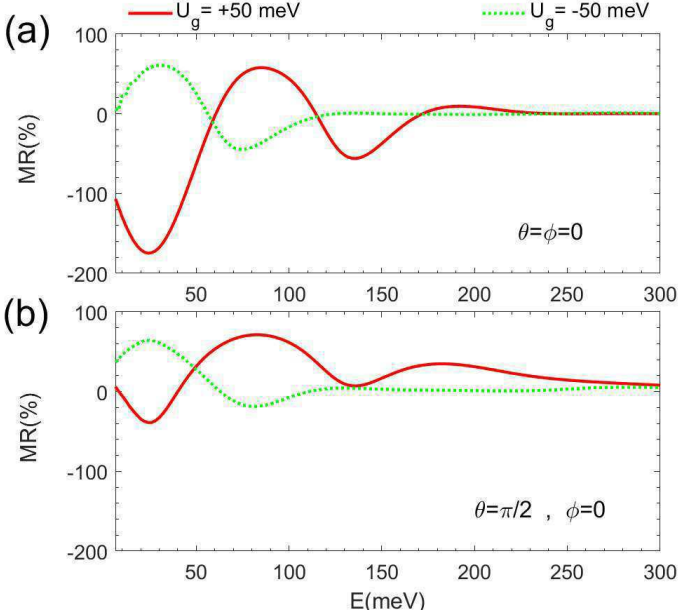}}
\caption{Calculated magnetoresistance ratio as a function of incident energy $E$ with $\Delta=40$\,meV and $d=W=4$\,nm.
The magnetization orientations align along the (a) $z$- and (b) $x$-axes.}
\label{F7}
\end{figure}

Energy dependence of the conductance at different $d$ and $W$ values with parallel magnetizations in the $z$- and $x$-direction is shown in Figs. 5(a) and (b), respectively. In the conductance spectra, one can see several resonant peaks whose positions are shifted by changing $d$ and $W$, simultaneously, regardless of the orientation of magnetizations. When the magnetizations align along the $z$-direction ($\theta=0, \phi=0$) the conductance around $E=U_g$ approaches to zero and a gap opens up in the conductance spectra as $d$ and $W$ are increased. This is a typical signature of broken time-reversal symmetry as a result of the normal component of $\bf m$ in the modulated regions. In contrast, the conductance spectra do not exhibit any gap opening when the magnetizations align in the $x$-direction ($\theta=\pi/2, \phi=0$). The conductance in the minimum case is greater than $0.2g_{0}$, indicating that due to the time reversal symmetry, the system is conductive for all surface electron energies, i.e.,  $E\leq U_{g}$ and  $E> U_{g}$.  This result clearly suggests that magnetization directions can control the electron charge transport on the surface of double-gated topological insulators. In addition, to examine the influence of warping effect on the electronic transport, we have depicted in the insets of Fig. 5 the conductance vs incident energy in the presence and absence of warping term (see Eq. (\ref{eq1})).  At low energies, the conductance is not affected by the warping strength, as expected from Fig. 1(b) and (c). At high energies relative to the Dirac point, however, the warping effect considerably enhances the conductance due to more transport channels resulting from the snowflake shape of the constant-energy contour.
  
\begin{figure}[h]
\centerline{\includegraphics[scale=0.55]{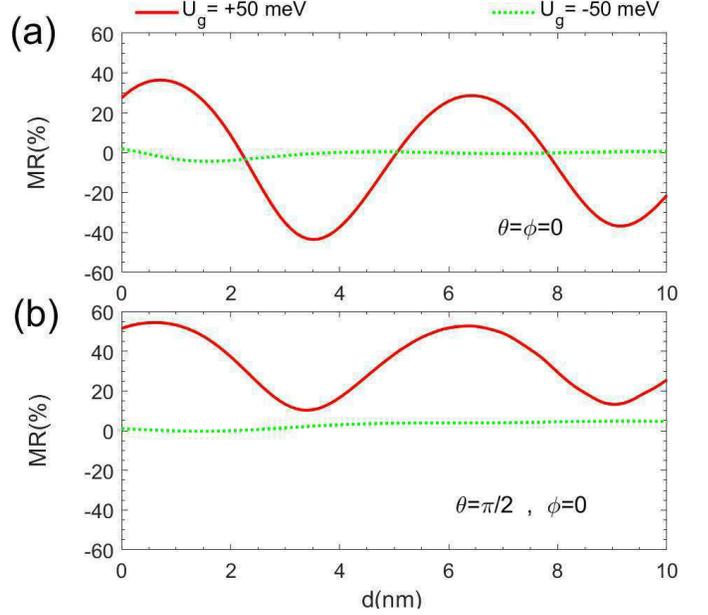}}
\caption{Calculated magnetoresistance ratio as a function of distance between the two magnetized regions at $E=150$\, meV with $\Delta=40$\,meV and $W=4$\,nm.
The magnetization orientations align along the (a) $z$- and (b) $x$-axes.}
\label{F8}
\end{figure}

The spin of topological surface states which lies in-plane is locked at right angles to the electron momentum, so that electrons in $\bf k$ and $-\bf k$ states carry opposite spins and possess opposite group velocities. Moreover, in the presence of hexagonal warping effect, an in-plane magnetic field not only deforms the constant-energy contour, but also shifts the Dirac point, as shown in Figs. 1(f) and (g) by a black dot. Therefore, the currents of electrons with opposite spin directions can no longer compensate each other, and hence, a spin-polarized current can be produced \cite{Li-NanoTec2014,Dankert-Nano-2015}. This property can be used to examine the magnetoresistance effect in our double-gated and magnetized structure. To do this, the electric conductance should be calculated in the two cases of parallel (Pa) and antiparallel (APa) magnetization directions. In Figs. 6(a) and (b), we have depicted the conductance versus energy, for both Pa and APa magnetization orientations along $z$- and $x$-axes, respectively. Also, we have shown the effect of positive and negative gate voltages on the conductance values for the two magnetization orientations. A positive (negative) gate voltage in the surface of topological insulator induces a double quantum barrier (well) structure and hence, the resonant states play the main role in the process of electron transport. For magnetizations in the $x$-direction, the conductance exhibits more oscillatory behaviour and the overlap between conductance values in the Pa and APa magnetizations is negligible compared to that for magnetizations in the $z$-direction, regardless of the sign of $U_g$.  The structure with quantum wells is more conductive than the structure with quantum barriers due to the electron tunnelling process, except for low energy electrons. Furthermore, due to the quantum interference of propagating waves in the region between the two magnetized segments, the conductance oscillations in the APa configuration are $180^{\circ}$ out-of-phase with respect to those in the Pa magnetic configuration, regardless of the sign of gate voltage. This indicates that one can obtain a large magnetoresistance effect in these double-modulated devices. 

The magnetoresistance ratio is defined as
\begin{equation}
\mathrm{MR}=\frac{G_{Pa}-G_{APa}}{G_{Pa}}\times 100\%\,
\end{equation}
where $G_{Pa}$ and $G_{APa}$ are the conductances in the parallel and antiparallel magnetization orientations, respectively.

In Figs. 7(a) and (b), we plot the magnetoresistance versus energy, when magnetization aligns along the $z$- and $x$-axes, respectively. We can see that MR exhibits an oscillatory behaviour due to the existence of resonant states in both double-barrier and double-well structures. In the case of magnetization in the $z$-axis, MR value can reach $\sim$ -170\% for the structure with double quantum barrier, whereas it reaches $\sim$ +60\% for double quantum well structure. On the other hand, when magnetization aligns along $x$-axis, MR oscillation decreases and the values are mostly positive. A positive magnetoresistance ratio which indicates a low resistance in the parallel magnetic configuration is of key importance for spintronic applications. For high energy electrons, MR approaches zero and the difference in the MR values in a double-barrier or double-well structure becomes negligible due to the suppression of conductance oscillations at high energies, as shown in Fig. 6.

Figure 8 shows the MR as a function of separation $d$ between the two magnetized regions. The MR exhibits large oscillations with distance $d$ in double-barrier structures, whereas the change in the magnetoresistance is very small in double-well structures. When $\bf m$ directions align along the $z$-axis, the MR finds positive and negative values by varying the separation distance. On the contrary, MR is fully positive and exceeds +50\%, when $\bf m$ orientations align in the $x$-direction. Note that MR is a function of energy as discussed in Fig. 7, nevertheless the nature of oscillations versus the separation distance between the two magnetized regions is independent of energy and it comes from spin-momentum locking of surface states. The MR oscillations closely resemble Ruderman-Kittel-Kasuya-Yoshida (RKKY) interactions between two impurity magnetic moments placed on the surface of a topological insulator and predict that the surface electronic states mediate such a RKKY interaction among the magnetic modulated regions \cite{Garate-Franz,Biswas-Balatsky}. It is worth mentioning that although magnetic impurities on the surface of topological insulators break time reversal symmetry by inducing a band gap in the energy dispersion, time reversal breaking does not form in a system consisting of magnetic modulated regions with magnetization directions along the $x$-axis.

\section{conclusions}
In summary, we have theoretically investigated the effects of double gate voltage and magnetism on Dirac fermions on the surface of a topological insulator.
Our findings demonstrate that the conductance decreases with increasing the exchange field and that the resonant states are not strongly affected by changing the magnetization orientation. The hexagonal warping effect increases the conductance at high energies when the constant-energy contour forms a snowflake shape. For magnetization direction along $x$-axis, time reversal symmetry is not broken and the system remains conductive for all surface electron energies.
Although the double-well structure is more conductive than the double-barrier structure, the MR oscillations corresponding to a positive gate voltage are more significant than those in the case of negative gate voltages. The MR is an oscillatory function in terms of energy and the separation distance between the two magnetized regions.
Our results suggest that the charge transport and MR effect on the surface of double-gated topological insulators can be effectively controlled by tuning the double gate voltage and magnetism.

\end{document}